\documentclass[12pt, centerh1]{article}
\usepackage{natbib}
\usepackage[margin=1in]{geometry}
\usepackage{amsmath,natbib,multirow}
\usepackage{amsfonts,mathrsfs,moreverb,lipsum}
\usepackage{graphicx,colonequals}
\usepackage{fixltx2e}
\usepackage{color}
\usepackage{caption}
\usepackage{subcaption}
\usepackage{mathtools}
\usepackage{pbox}
\newcommand{\btheta}{{\boldsymbol{\theta}}}
\newcommand{\bvtheta}{{\boldsymbol{\vartheta}}}
\newcommand{\pig}{\pi_g}

\newcommand{\bLambda}{{\boldsymbol{\Lambda}}}

\newcommand{\bfx}{{\bf x}}

\newcommand{\bfz}{{\bf z}}

\newcommand{\vecmu}{\mbox{\boldmath$\mu$}}

\newcommand{\vecx}{\mathbf{x}}

\newcommand{\vecX}{\mathbf{X}}
\newcommand{\vecU}{\mathbf{U}}

\newcommand{\loada}{\mathbf\Lambda}
\newcommand{\loadb}{\mathbf\Delta}
\newcommand{\matsig}{\mathbf\Sigma}
\newcommand{\matPsi}{\mathbf\Psi}

\newcommand{\diag}{\,\mbox{diag}}
\newcommand{\tr}{\,\mbox{tr}}

\newcommand{\vecA}{\mathbf{A}}
\newcommand{\vecC}{\mathbf{C}}
\newcommand{\vecL}{\mathbf{L}}
\newcommand{\vecB}{\mathbf{B}}

\newcommand{\vecS}{\mathbf{S}}

\newcommand{\matd}{\mathbf{D}}

\newcommand{\matm}{\mathbf{M}}

\newcommand{\ident}{\mathbf{I}}

\newcommand{\vecepsilon}{\mbox{\boldmath$\varepsilon$}}

\newcommand{\vecc}{\text{vec}}
\newcommand{\fX}{\mathscr{X}}
\newcommand{\fV}{\mathscr{V}}
\newcommand{\fU}{\mathscr{U}}
\newcommand{\fE}{\mathscr{E}}
\newcommand{\fY}{\mathscr{Y}}
\newcommand{\mathn}{\mathcal{N}}
\newcommand{\tgamma}{\kappa}
\newcommand{\bzero}{{\bf 0}}
\newcommand{\fR}{\mathscr{R}}

\newcommand{\matsigs}{\matsig_g^*}
\newcommand{\matsigsih}{(\hat{\matsig}_g^*)^{-1}}
\newcommand{\matsigsi}{(\matsig_g^*)^{-1}}
\allowdisplaybreaks
\title{Mixtures of Skewed Matrix Variate\\ Bilinear Factor Analyzers}
\author{Michael P.B.\ Gallaugher and Paul D.\ McNicholas}
\date{\small Dept.\ of Mathematics \& Statistics, McMaster University, Hamilton, Ontario, Canada.}
\begin{document}
\maketitle{}

\begin{abstract}
In recent years, data have become increasingly higher dimensional and, therefore, an increased need has arisen for dimension reduction techniques for clustering. Although such techniques are firmly established in the literature for multivariate data, there is a relative paucity in the area of matrix variate, or three-way, data. Furthermore, the few methods that are available all assume matrix variate normality, which is not always sensible if cluster skewness or excess kurtosis is present. Mixtures of bilinear factor analyzers using skewed matrix variate distributions are proposed. In all, four such mixture models are presented, based on matrix variate skew-$t$, generalized hyperbolic, variance-gamma, and normal inverse Gaussian distributions, respectively.
\\[-10pt]

\end{abstract}

\section{Introduction}
Classification is the process of finding and analyzing underlying group structure in heterogenous data. This problem can be framed as the search for class labels of unlabelled observations. In general, some (non-trivial) proportion of observations have known labels. A special case of classification, known as clustering, occurs when none of the observations have known labels. One common approach for clustering is mixture model-based clustering, which makes use of a finite mixture model. In general, a $G$-component finite mixture model assumes that a multivariate random variable $\vecX$ has density
\begin{equation}\label{eqn:lll}
f(\vecx~|~\bvtheta)=\sum_{g=1}^G\pig f_g(\vecx~|~\btheta_g),
\end{equation}
where $\bvtheta=\left(\pi_1,\pi_2,\ldots,\pi_G,\btheta_1,\btheta_2,\ldots,\btheta_G\right)$, $f_g(\cdot)$ is the $g$th component density, and $\pig>0$ is the $g$th mixing proportion such that $\sum_{i=1}^G\pig=1$. Note that the notation used in \eqref{eqn:lll} corresponds to the multivariate case and, save for Appendix~A, $\fX$ will hereafter represent a matrix variate random variable with $\vecX$ denoting its realization.

\cite{mcnicholas16a} traces the relationship between mixture models and clustering to \cite{tiedeman55}, who uses a component of a mixture model to define a cluster. The mixture model was first used for clustering by \cite{wolfe65} who considered a Gaussian mixture model. Other early uses of Gaussian mixture models for clustering can be found in \cite{baum70} and \cite{scott71}. 
Although the Gaussian mixture model is attractive due to its mathematical properties, it can be problematic when dealing with outliers and asymmetry in clusters and thus there has been an interest in non-Gaussian mixtures in the multivariate setting. Some examples of mixtures of symmetric distributions that parameterize tail weight include the $t$ distribution \citep{peel00,andrews11a,andrews12,lin14} and the power exponential distribution \citep{dang15}. There has also been work in the area of skewed distributions such as the skew-$t$ distribution, \citep{lin10,vrbik12,vrbik14,lee14,murray14b,murray14a}, the normal-inverse Gaussian (NIG) distribution \citep{karlis09}, the shifted asymmetric Laplace (SAL) distribution \citep{morris13b,franczak14}, the variance-gamma distribution \citep{smcnicholas17}, the generalized hyperbolic distribution \citep{browne15}, the hidden truncation hyperbolic distribution \citep{murray17b}, the joint generalized hyperbolic distribution \citep{tang18}, and the coalesced generalized hyperbolic distribution \citep{tortora19}.

There has also been an increased interest in model-based clustering of matrix variate, or three-way, data such as multivariate longitudinal data and images. Examples include the work of \cite{viroli11a} and \cite{Anderlucci15}, who consider mixtures of matrix variate normal distributions for clustering. \cite{viroli11b} further considers model-based clustering with matrix variate normal distributions in the Bayesian paradigm.  More recently, \cite{gallaugher18a} investigate mixtures of four skewed matrix distributions--- namely, the matrix variate skew-$t$, generalized hyperbolic, variance-gamma, and normal inverse Gaussian (NIG) distributions \citep[see][for details about these matrix variate distributions]{gallaugher17b}---and consider classification of greyscale Arabic numerals. \cite{melnykov18,melnykov19} consider modelling skewness by means of transformations.

The main problem with all of the aforementioned methods, for both the multivariate and matrix variate cases, arises when the dimensionality of the data increases. Although the problem of dealing with high-dimensional data has been thoroughly addressed in the case of multivariate data, there is relative paucity of work for matrix variate data. In the matrix variate case, matrix variate bilinear probabilistic principal component analysis was developed by \cite{zhao12}. More recently, \cite{gallaugher18b} considered the closely-related mixture of matrix variate bilinear factor analyzers (MMVBFA) model for clustering. The MMVBFA model can be viewed as a matrix variate generalization of the mixture of factor analyzers model \citep{ghahramani97} in the multivariate case. Although these methods allow for simultaneous dimension reduction and clustering, they both assume matrix variate normality which is not sensible if cluster skewness or heavy tails are present. Herein we present an extension of the MMVBFA model to skewed distributions; specifically, the matrix variate skew-$t$, generalized hyperbolic, variance-gamma, and NIG distributions.

\section{Background}
\subsection{Generalized Inverse Gaussian Distribution} 
The generalized inverse Gaussian distribution has two different parameterizations, both of which will be utilized herein. A random variable $Y$ has a generalized inverse Gaussian (GIG) distribution parameterized by $a, b$ and $\lambda$, and denoted $\text{GIG}(a,b,\lambda)$, if its probability density function can be written as
$$
f(y|a, b, \lambda)=\frac{\left({a}/{b}\right)^{\frac{\lambda}{2}}y^{\lambda-1}}{2K_{\lambda}(\sqrt{ab})}\exp\left\{-\frac{ay+{b}/{y}}{2}\right\},
$$
for $y>0$, $a,b\in\mathbb{R}^+$ and $\lambda\in\mathbb{R}$, where

$$
K_{\lambda}(u)=\frac{1}{2}\int_{0}^{\infty}y^{\lambda-1}\exp\left\{-\frac{u}{2}\left(y+\frac{1}{y}\right)\right\}dy
$$
is the modified Bessel function of the third kind with index $\lambda$. 
Expectations of some functions of a GIG random variable have a mathematically tractable form, e.g.:
\begin{equation}
\mathbb{E}(Y)=\sqrt{\frac{b}{a}}\frac{K_{\lambda+1}(\sqrt{ab})}{K_{\lambda}(\sqrt{ab})},
\label{eq:ai}
\end{equation}
\begin{equation}
\mathbb{E}\left({1}/{Y}\right)=\sqrt{\frac{a}{b}}\frac{K_{\lambda+1}(\sqrt{ab})}{K_{\lambda}(\sqrt{ab})}-\frac{2\lambda}{b},
\label{eq:bi}
\end{equation}
\begin{equation}
\mathbb{E}(\log Y)=\log\left(\sqrt{\frac{b}{a}}\right)+\frac{1}{K_{\lambda}(\sqrt{ab})}\frac{\partial}{\partial \lambda}K_{\lambda}(\sqrt{ab}).
\label{eq:ci}
\end{equation}

Although this parameterization of the GIG distribution will be useful for parameter estimation, the alternative parameterization given by
\begin{equation}
g(y|\omega,\eta,\lambda)= \frac{\left({w}/{\eta}\right)^{\lambda-1}}{2\eta K_{\lambda}(\omega)}\exp\left\{-\frac{\omega}{2}\left(\frac{w}{\eta}+\frac{\eta}{w}\right)\right\},
\label{eq:I}
\end{equation}
where $\omega=\sqrt{ab}$ and $\eta=\sqrt{a/b}$, is used when deriving the generalized hyperbolic distribution \citep[see][]{browne15}. For notational clarity, we will denote the parameterization given in \eqref{eq:I} by $\text{I}(\omega,\eta,\lambda)$.

\subsection{Matrix Variate Distributions}
As in the multivariate case, the most mathematically tractable matrix variate distribution is the matrix variate normal. An $n\times p$ random matrix $\fX$ follows an $n\times p$ matrix variate normal distribution with $n\times p$ location matrix $\matm$ and scale matrices $\matsig$ and $\matPsi$, of dimensions $n\times n$ and $p\times p$, respectively, denoted $\mathcal{N}_{n\times p}(\matm, \matsig, \matPsi)$, if the density of $\fX$ is
\begin{equation}
f(\vecX~|~\matm, \matsig, \matPsi)=\frac{1}{(2\pi)^{\frac{np}{2}}|\matsig|^{\frac{p}{2}}|\matPsi|^{\frac{n}{2}}}\exp\left\{-\frac{1}{2}\tr\left(\matsig^{-1}(\vecX-\matm)\matPsi^{-1}(\vecX-\matm)'\right)\right\}.
\end{equation}
The matrix variate normal distribution is related to the multivariate normal distribution, as discussed in \cite{harrar08}, via
$\fX\sim \mathcal{N}_{n\times p}(\matm,\matsig,\matPsi) \iff \vecc(\fX)\sim \mathcal{N}_{np}(\vecc(\matm),\matPsi\otimes \matsig),$
where $\mathcal{N}_{np}(\cdot)$ is the multivariate normal density with dimension $np$, $\vecc(\matm)$ is the vectorization of $\matm$, and $\otimes$ denotes the Kronecker product.
Although the matrix variate normal distribution is popular, there are other well-known examples of matrix variate distributions. For example, the Wishart distribution \citep{Wishart} is the distribution of the sample covariance matrix in the multivariate normal case. There are also a few formulations of a matrix variate skew normal distribution \citep{chen2005,dominguez2007,harrar08}.

More recently, \cite{gallaugher17a,gallaugher17b} derived a total of four skewed matrix variate distributions using a variance-mean matrix variate mixture approach. This assumes that a random matrix $\fX$ can be written as 
\begin{equation}
\fX=\matm+W\vecA+\sqrt{W}\fV,
\label{eq:mvmix}
\end{equation}
where $\matm$ and $\vecA$ are $n\times p$ matrices representing the location and skewness, respectively, $\fV \sim \mathcal{N}_{n \times p}\left(\bf{0}, \matsig , \matPsi \right)$, and $W>0$ is a random variable with density $h(w|\btheta)$. 
\cite{gallaugher17a}, show that the matrix variate skew-$t$ distribution, with $\nu$ degrees of freedom, arises from \eqref{eq:mvmix} with $W^{\text{ST}}\sim \text{IGamma}(\nu/2,\nu/2)$, where $\text{IGamma}(\cdot)$ denotes the inverse-gamma distribution with density 
$$
f(y~|~a,b)=\frac{b^a}{\Gamma(a)}y^{-a-1}\exp\left\{-\frac{b}{y}\right\},
$$
for $y>0$ and $a,b \in \mathbb{R}^+$.
The resulting density of $\fX$ is 
\begin{align*}
f_{\text{\tiny MVST}}(\vecX~|~\bvtheta)=&\frac{2\left(\frac{\nu}{2}\right)^{\frac{\nu}{2}}\exp\left\{\tr(\matsig^{-1}(\vecX-\matm)\matPsi^{-1}\vecA') \right\} }{(2\pi)^{\frac{np}{2}}| \matsig |^{\frac{p}{2}} |\matPsi |^{\frac{n}{2}}\Gamma(\frac{\nu}{2})}  \left(\frac{\delta(\vecX;\matm,\matsig,\matPsi)+\nu}{\rho(\vecA,\matsig,\matPsi)}\right)^{-\frac{\nu+np}{4}} \\ & \qquad\qquad\qquad\qquad\times
 K_{-\frac{\nu+np}{2}}\left(\sqrt{\left[\rho(\vecA,\matsig,\matPsi)\right]\left[\delta(\vecX;\matm,\matsig,\matPsi)+\nu\right]}\right),
\end{align*}
where 
$
\delta(\vecX;\matm,\matsig,\matPsi)=\tr(\matsig^{-1}(\vecX-\matm)\matPsi^{-1}(\vecX-\matm)')$, $\rho(\vecA,\matsig,\matPsi)=\tr(\matsig^{-1}\vecA\matPsi^{-1}\vecA')
$
and $\nu>0$.
For notational clarity, this distribution will be denoted by $\text{MVST}(\matm,\vecA,\matsig,\matPsi,\nu)$.

In \cite{gallaugher17b}, one of the distributions considered is a matrix variate generalized hyperbolic distribution. This again arises from \eqref{eq:mvmix} with $W^{\text{GH}}\sim\text{I}(\omega,1,\lambda)$. This distribution will be denoted by $\text{MVGH}(\matm,\vecA,\matsig,\matPsi,\lambda,\omega)$, and the density is 
\begin{align*}
f_{\text{\tiny MVGH}}(\vecX|\bvtheta)=&\frac{\exp\left\{\tr(\matsig^{-1}(\vecX-\matm)\matPsi^{-1}\vecA') \right\}}{(2\pi)^{\frac{np}{2}}| \matsig |^{\frac{p}{2}} |\matPsi |^{\frac{n}{2}}K_{\lambda}(\omega)}  \left(\frac{\delta(\vecX;\matm,\matsig,\matPsi)+\omega}{\rho(\vecA,\matsig,\matPsi)+\omega}\right)^{\frac{\left(\lambda-\frac{np}{2}\right)}{2}} \\ & \times
 K_{\left(\lambda-{np}/{2}\right)}\left(\sqrt{\left[\rho(\vecA,\matsig,\matPsi)+\omega\right]\left[\delta(\vecX;\matm,\matsig,\matPsi)+\omega\right]}\right),
\end{align*}
where $\lambda\in \mathbb{R}$ and $\omega\in\mathbb{R}^+$.

The matrix variate variance-gamma distribution, also derived in \cite{gallaugher17b} and denoted $\text{MVVG}(\matm,\vecA,\matsig,\matPsi,\gamma)$, arises from \eqref{eq:mvmix} with $W^{\text{VG}}\sim\text{gamma}(\gamma,\gamma)$, where $\text{gamma}(\cdot)$ denotes the gamma distribution with density 
$$
f(y~|~a,b)=\frac{b^a}{\Gamma(a)}y^{a-1}\exp\left\{-by\right\},
$$
for $y>0$ and $a,b\in\mathbb{R}^+$
The density of the random matrix $\fX$ with this distribution is
\begin{align*}
f_{\text{\tiny MVVG}}(\vecX|\bvtheta)=&\frac{2\gamma^{\gamma}\exp\left\{\tr(\matsig^{-1}(\vecX-\matm)\matPsi^{-1}\vecA') \right\}}{(2\pi)^{\frac{np}{2}}| \matsig |^{\frac{p}{2}} |\matPsi |^{\frac{n}{2}}\Gamma(\gamma)}  \left(\frac{\delta(\vecX;\matm,\matsig,\matPsi)}{\rho(\vecA,\matsig,\matPsi)+2\gamma}\right)^{\frac{\left(\gamma-{np}/{2}\right)}{2}} \\
&\times  K_{\left(\gamma-\frac{np}{2}\right)}\left(\sqrt{\left[\rho(\vecA,\matsig,\matPsi)+2\gamma\right]\left[\delta(\vecX;\matm,\matsig,\matPsi)\right]}\right),
\end{align*}
where $\gamma>0$.

Finally, the matrix variate NIG distribution arises when $W^{\text{NIG}}\sim\text{IG}(1,\kappa)$, where $\text{IG}(\cdot)$ denotes the inverse-Gaussian distribution with density
$$
f(y~|~\delta,\kappa)=\frac{\delta}{\sqrt{2\pi}}\exp\{\delta\kappa\}y^{-\frac{3}{2}}\exp\left\{-\frac{1}{2}\left(\frac{\delta^2}{y}+\kappa^2y\right)\right\},
$$
for $y>0$, $\delta,\kappa\in\mathbb{R}^+$.
The density of $\fX$ is
\begin{align*}
f_{\text{\tiny MVNIG}}(\vecX|\bvtheta)&=\frac{2\exp\left\{\tr(\matsig^{-1}(\vecX-\matm)\matPsi^{-1}\vecA')+\tgamma\right\}
}{(2\pi)^{\frac{np+1}{2}}| \matsig |^{\frac{p}{2}} |\matPsi |^{\frac{n}{2}}}\left(\frac{\delta(\vecX;\matm,\matsig,\matPsi)+1}{\rho(\vecA,\matsig,\matPsi)+\tgamma^2}\right)^{-{\left(1+np\right)}/{4}}\\
&\times K_{-{(1+np)}/{2}}\left(\sqrt{\left[\rho(\vecA,\matsig,\matPsi)+\tgamma^2\right]\left[\delta(\vecX;\matm,\matsig,\matPsi)+1\right]}\right),
\end{align*}
where $\tgamma>0$. This distribution is denoted by $\text{MVNIG}(\matm,\vecA,\matsig,\matPsi,\tgamma)$.

\subsection{Matrix Variate Factor Analysis}
Readers who may benefit from the context provided by the mixture of factor analyzers model should consult Appendix A. \cite{xie08} and \cite{yu08} consider a matrix variate extension of probabilistic principal components analysis (PPCA), which assumes an $n\times p$ random matrix $\fX$ can be written
\begin{equation}
\fX=\matm+\loada\fU\loadb'+\fE,
\label{eq:MVPPCA}
\end{equation}
where $\matm$ is an $n\times p$ location matrix, $\loada$ is an $n\times q$ matrix of column factor loadings, $\loadb$ is a $p\times r$ matrix of row factor loadings, $\fU\sim \mathcal{N}_{q\times r}({\bf 0},\ident_q,\ident_r)$, and $\fE\sim\mathcal{N}_{n\times p}({\bf 0},\sigma^2\ident_n,\sigma^2\ident_p)$. It is assumed that $\fU$ and $\fE$ are independent of each other. The main disadvantage of this model is that, in general, $\fX$ does not follow a matrix variate normal distribution. 

\cite{zhao12} present bilinear probabilistic principal component analysis (BPPCA), which extends \eqref{eq:MVPPCA} by adding two projected error terms. The resulting model assumes $\fX$ can be written
$\fX=\matm+\loada\fU\loadb'+\loada\fE^B+\fE^A\loadb'+\fE,$
where $\fU$ and $\fE$ are defined as in \eqref{eq:MVPPCA}, $\fE^B\sim \mathcal{N}_{q\times p}({\bf 0},\ident_q,\sigma_B\ident_p)$, and $\fE^A\sim \mathcal{N}_{n\times r}({\bf 0}, \sigma_A\ident_n,\ident_r)$. In this model, it is assumed that $\fU, \fE^B, \fE^A$, and $\fE$ are all independent of each other.
\cite{gallaugher18b} further extend this to matrix variate factor analysis and consider a mixture of matrix variate bilinear factor analyzers (MMVBFA) model. For MMVBFA, \cite{gallaugher18b} generalize BPPCA by removing the isotropic constraints so that $\fE^B\sim \mathcal{N}_{q\times p}({\bf 0},\ident_q,\matPsi)$, $\fE^A\sim \mathcal{N}_{n\times r}({\bf 0},\matsig,\ident_r)$, and $\fE\sim\mathcal{N}_{n\times p}({\bf 0},\matsig,\matPsi)$, where $\matsig=\diag\{\sigma_1,\sigma_2,\ldots,\sigma_n\}$ with $\sigma_i>0$, and $\matPsi=\diag\{\psi_1,\psi_2, \ldots, \psi_p\}$ with $\psi_i>0$. With these slight modifications, it can be shown that
$\fX\sim\mathcal{N}_{n\times p}(\matm,\loada\loada'+\matsig,\loadb\loadb'+\matPsi)$, similarly to its multivariate counterpart (Appendix~A).

It is important to note that the term ``column factors" refers to reduction in the dimension of the columns, which is equivalent to the number of rows, and not a reduction in the number of columns. Likewise, the term ``row factors" refers to the reduction in the dimension of the rows (number of columns). As discussed by \cite{zhao12}, the interpretation of the terms $\fE^B$ and $\fE^A$ are the row and column noise, respectively, whereas the term $\fE$ is the common noise.

\section{Mixture of Skewed Matrix Variate Bilinear Factor Analyzers}
\subsection{Model Specification}
We now consider a mixture of skewed bilinear factor analyzers according to one of the four skewed distributions discussed previously. Each random matrix $\fX_i$ from a random sample distributed according to one of the four distributions can be written as
$$
\fX_i=\matm_g+W_{ig}\vecA_g+\fV_{ig}
$$
with probability $\pi_g$ for $g\in\{1,2,\ldots,G\}$, $\pig>0$, $\sum_{i=1}^G\pig=1$, where $\matm_g$ is the location of the $g$th component, $\vecA_g$ is the skewness, and $W_{ig}$ is a random variable with density $h(w_{ig}|\btheta_g)$. The distribution of the random variable $W_{ig}$---and so the density $h(w_{ig}|\btheta_g)$---will change depending on the distribution of $\fX_i$, i.e., skew-$t$, generalized hyperbolic, variance-gamma, or NIG. Assume also that $\fV_{ig}$ can be written as
$$
\fV_{ig}=\loada_g\fU_{ig}\loadb_g'+\loada_g\fE_{ig}^B+\fE_{ig}^A\loadb_g'+\fE_{ig},
$$
where $\loada_g$ is a $n\times q$ matrix of column factor loadings, $\loadb_g$ is a $p\times r$ matrix of row factor loadings, and 
\begin{align*}
\fU_{ig}|w_{ig}&\sim\mathn_{q\times r}(\bzero,w_{ig}\ident_q,\ident_p),
&\fE_{ig}^B|w_{ig}\sim\mathn_{q\times p}(\bzero,w_{ig}\ident_q,\matPsi_g),\\
\fE_{ig}^A|w_{ig}&\sim\mathn_{n\times r}(\bzero,w_{ig}\matsig_g,\ident_r),
&\fE_{ig}|w_{ig}\sim\mathn_{n\times p}(\bzero,w_{ig}\matsig_g,\matPsi_g).
\end{align*}
Note that $\fU_{ig},\fE_{ig}^B,\fE_{ig}^A$, and $\fE_{ig}$ are all independently distributed and independent of each other.

To facilitate clustering, introduce the indicator $z_{ig}$, where $z_{ig}=1$ if observation $i$ belongs to group $g$, and $z_{ig}=0$ otherwise. Then, it can be shown that
$$
\fX_i~|~z_{ig}=1\sim\text{D}_{n\times p}(\matm_g,\vecA_g,\matsig_g+\loada_g\loada_g',\matPsi_g+\loadb_g\loadb_g',\btheta_g),
$$
where $\text{D}$ is the distribution in question, and $\btheta_g$ is the set of parameters related to the distribution of $W_{ig}$.

As in the matrix variate normal case, this model has a two stage interpretation given by
\begin{equation*}\begin{split}
\fX_i&=\matm_g+W_{ig}\vecA+\loada_g\fY_{ig}^B+\fR_{ig}^B,\\
\fY_{ig}^B&=\fU_{ig}\loadb_g'+\fE_{ig}^B,\\
\fR_{ig}^B&=\fE_{ig}^A\loadb_{g}'+\fE_{ig},
\end{split}\end{equation*}
and 
\begin{equation*}\begin{split}
\fX_i&=\matm_g+W_{ig}\vecA+\fY_{ig}^A\loadb_g'+\fR_{ig}^A,\\
\fY_{ig}^A&=\loada_g\fU_{ig}+\fE_{ig}^A,\\
\fR_{ig}^A&=\loada_g\fE_{ig}^B+\fE_{ig},
\end{split}\end{equation*}
which will be useful for parameter estimation.

\subsection{Parameter Estimation}
Suppose we observe the $N$ $n\times p$ matrices $\vecX_1,\ldots,\vecX_N$ distributed according to one of the four distributions. We assume that these data are incomplete and employ an alternating expectation conditional maximization (AECM) algorithm \citep{meng97}. This algorithm is now described after initialization.

\paragraph{AECM 1st Stage} The complete-data in the first stage consists of the observed data $\vecX_i$, the latent variables ${\bf W}_i=(W_{i1},\ldots,W_{iG})'$, and the unknown group labels $\bfz_{i}=(z_{i1},\ldots,z_{iG})'$ for $i=1,2,\ldots,N$. In this case, the complete-data log-likelihood is
\begin{equation*}\begin{split}
\ell_{\text{C1}}=C+&\sum_{i=1}^N\sum_{g=1}^Gz_{ig}\bigg[\log \pig+\log h(w_{ig}|\btheta_g)-\frac{1}{2}\tr\bigg\{\frac{1}{W_{ig}}(\matsig_g^*)^{-1}(\vecX_i-\matm_g)(\matPsi_g^*)^{-1}(\vecX_i-\matm_g)'\\
&-(\matsig_g^*)^{-1}(\vecX_i-\matm_g)(\matPsi_g^*)^{-1}\vecA_g'-(\matsig_g^*)^{-1}\vecA_g(\matPsi_g^*)^{-1}(\vecX_i-\matm_g)'+W_{ig}(\matsig_g^*)^{-1}\vecA_g(\matPsi_g^*)^{-1}\vecA_g'\bigg\}\bigg],
\end{split}\end{equation*}
where $\matsig_g^{*}=\matsig_g+\loada_g\loada_g'$, $\matPsi_g^*=\matPsi_g+\loadb_g\loadb_g'$ and $C$ is constant with respect to the parameters.

In the E-step, we calculate the following conditional expectations:
\begin{equation*}\begin{split}
\hat{z}_{ig}&=\frac{\pig f(\vecX_i~|~\hat{\bvtheta}_g)}
{\sum_{h=1}^G\pi_h f(\vecX_i~|~\hat{\bvtheta}_h)},\qquad\qquad
a_{ig}=\mathbb{E}(W_{ig}~|~\vecX_i,z_{ig}=1,\hat{\bvtheta}_g),\\
b_{ig}&=\mathbb{E}\left(\frac{1}{W_{ig}}~\bigg|~\vecX_i,z_{ig}=1,\hat{\bvtheta}_g\right),\qquad
c_{ig}=\mathbb{E}(\log W_{ig}~|~\vecX_i,z_{ig}=1,\hat{\bvtheta}_g).\\
\end{split}\end{equation*}
As usual, all expectations are conditional on current parameter estimates; however, to avoid cluttered notation, we do not use iteration-specific notation. Although these expectations are dependent on the distribution in question, it can be shown that
\begin{align*}
W_{ig}^{\text{ST}}~|~\vecX_i, z_{ig}=1&\sim \text{GIG}\left(\rho(\vecA_g,\matsig^*_g,\matPsi^*_g),\delta(\vecX;\matm_g,\matsig^*_g,\matPsi^*_g)+\nu_g,-(\nu_g+np)/2\right),\\
W_{ig}^{\text{GH}}~|~\vecX_i, z_{ig}=1&\sim \text{GIG}\left(\rho(\vecA_g,\matsig^*_g,\matPsi^*_g)+\omega_g,\delta(\vecX;\matm_g,\matsig^*_g,\matPsi^*_g)+\omega_g,\lambda_g-{np}/{2}\right),\\
W_{ig}^{\text{VG}}~|~\vecX_i, z_{ig}=1&\sim \text{GIG}\left(\rho(\vecA_g,\matsig^*_g,\matPsi^*_g)+2\gamma_g,\delta(\vecX;\matm_g,\matsig^*_g,\matPsi^*_g),\gamma_g-{np}/{2}\right),\\
W_{ig}^{\text{NIG}}~|~\vecX_i, z_{ig}=1&\sim \text{GIG}\left(\rho(\vecA_g,\matsig^*_g,\matPsi^*_g)+\tgamma_g^2,\delta(\vecX;\matm_g,\matsig^*_g,\matPsi^*_g)+1,-{(1+np)}/{2}\right).
\end{align*}
Therefore, the exact updates are obtained by using the expectations given in \eqref{eq:ai}--\eqref{eq:ci} for appropriate values of $\lambda, a$, and $b$. 

In the M-step, we update $\hat{\pi}_g,\hat\matm_g,\hat\vecA_g$, and $\hat\btheta_g$ for $g=1,\ldots,G$. We have:
$$
\hat{\pi}_g=\frac{N_g}{N}, \qquad 
\hat{\matm}_g=\frac{\sum_{i=1}^N\hat{z}_{ig}\left(\overline{a}_gb_{ig}-1\right)\vecX_i}{\sum_{i=1}^N\hat{z}_{ig}\overline{a}_gb_{ig}-N_g}, \qquad 
\hat{\vecA}=\frac{\sum_{i=1}^N\hat{z}_{ig}\left(\overline{b}_g-b_{ig}\right)\vecX_i}{\sum_{i=1}^N\hat{z}_{ig}\overline{a}_gb_{ig}-N_g}, 
$$
where 
$$
N_g=\sum_{i=1}^N\hat{z}_{ig},\qquad\overline{a}_g=\frac{1}{N_g}{\sum_{i=1}^N\hat{z}_{ig}a_{ig}},\qquad \overline{b}_g=\frac{1}{N_g}{\sum_{i=1}^N\hat{z}_{ig}b_{ig}}.
$$
The update for $\btheta_g$ is dependent on the distribution and will be identical to one of those given in \cite{gallaugher18a}.

\paragraph{AECM Stage 2} In the second stage, the complete-data consists of the observed data $\vecX_i$, the latent variables ${\bf W}_i$, the unknown group labels $\bfz_{i}$, and the latent matrices ${\boldsymbol \fY}^B_i=(\fY^B_{i1},\ldots,\fY^B_{iG})$ for $i=1,\ldots,N$. The complete-data log-likelihood at this stage is
\begin{align*}
\ell_{\text{C}2}&=C+\sum_{i=1}^N\sum_{g=1}^Gz_{ig}\big[\log \pig+\log h(W_{ig}|\nu_g)+\log \phi_{q\times p}(\fY^B_{ig}|\bzero,W_{ig}\ident_q,\matPsi^{*}_g)\\&
\quad+\log \phi_{n\times p}(\vecX_i|\matm_g+W_{ig}\vecA_g+\loada_g\fY_{ig}^B,W_{ig}\matsig_g,\matPsi_g^*)\big]\\
&=C+\sum_{i=1}^N\sum_{g=1}^G-\frac{1}{2}z_{ig}\bigg[-p\log|\matsig_g|+\tr\bigg\{\frac{1}{W_{ig}}\matsig_g^{-1}(\vecX_i-\matm_g)(\matPsi_g^*)^{-1}(\vecX_i-\matm_g)'\\
&\quad-\matsig_g^{-1}(\vecX_i-\matm_g)(\matPsi_g^*)^{-1}\vecA_g'\bigg\}
-\frac{1}{W_{ig}}\matsig_g^{-1}(\vecX_i-\matm_g)(\matPsi_g^*)^{-1}{\fY_{ig}^B}'\loada_g'-\matsig_g^{-1}\vecA_g(\matPsi_g^*)^{-1}(\vecX_i-\matm_g)'\\
&\quad+W_{ig}\matsig_g^{-1}\vecA_g(\matPsi_g^*)^{-1}\vecA_g'+\matsig_g^{-1}\vecA_g(\matPsi_g^*)^{-1}{\fY_{ig}^B}'\loada_g'
-\frac{1}{W_{ig}}\matsig_g^{-1}\loada_g\fY_{ig}^B(\matPsi_g^*)^{-1}(\vecX_i-\matm_g)'\\
&\quad+\matsig_g^{-1}\loada_g\fY_{ig}^B(\matPsi_g^*)^{-1}\vecA_g'+\frac{1}{W_{ig}}\matsig_g^{-1}\loada_g{\fY_{ig}^B}(\matPsi_g^*)^{-1}{\fY^B_{ig}}'\loada_g'\bigg].
\end{align*}

In the E-step, it can be shown that
\begin{equation*}\begin{split}
\fY_{ig}^B~|~\vecX_i,&W_{ig},z_{ig}=1\sim\\& \mathn_{q\times p}((\ident_q+\loada_g'\matsig_g^{-1}\loada_g)^{-1}\loada_g'\matsig_g^{-1}(\vecX_i-\matm_g-W_{ig}\vecA_g),W_{ig}(\ident_q+\loada_g'\matsig_g^{-1}\loada_g)^{-1},\matPsi^*_g)
\end{split}\end{equation*}
and so we can calculate the expectations
\begin{equation*}\begin{split}
{\bf E}_{1ig}^{(2)}&\coloneqq\mathbb{E}[\fY_{ig}^B|\hat{\bvtheta},\vecX_i,z_{ig}=1]
=\vecL_g(\vecX_i-\hat{\matm}_g-a_{ig}\hat{\vecA}_g),\\
{\bf E}_{2ig}^{(2)}&\coloneqq\mathbb{E}\left[\frac{1}{W_{ig}}\fY_{ig}^B\bigg|\hat{\bvtheta},\vecX_i,z_{ig}=1\right]
=\vecL_g(b_{ig}(\vecX_i-\hat{\matm}_g)-\hat{\vecA}_g),\\
{\bf E}_{3ig}^{(2)}&\coloneqq\mathbb{E}\left[\frac{1}{W_{ig}}\fY_{ig}^B(\matPsi_g^*)^{-1}{\fY_{ig}^B}'\bigg|\hat{\bvtheta},\vecX_i,z_{ig}=1\right]\\
&=p(\ident_q+\hat{\loada}_g'\hat{\matsig}_g^{-1}\hat{\loada}_g)^{-1}+b_{ig}\vecL_g(\vecX_i-\hat{\matm}_g)(\matPsi_g^*)^{-1}(\vecX_i-\hat{\matm}_g)'\vecL_g'\\
&\quad-\vecL_g((\vecX_i-\hat{\matm}_g)(\hat{\matPsi}_g^*)^{-1}\hat{\vecA}_g'+\hat{\vecA}_g(\hat{\matPsi}_g^*)^{-1}(\vecX_i-\hat{\matm}_g)')\vecL_g'+a_{ig}\vecL_g\hat{\vecA}_g(\hat{\matPsi}_g^*)^{-1}\hat{\vecA}_g'\vecL_g',
\end{split}\end{equation*}
where $\vecL_g=(\ident_q+\hat{\loada}_g'\hat{\matsig}_g^{-1}\hat{\loada}_g)^{-1}\hat{\loada}_g'\hat{\matsig}_g^{-1}$.

In the M-step, the updates for $\loada_g$ and $\matsig_g$ are calculated. These updates are given by
$$
\hat{\loada}_g=\sum_{i=1}^N\hat{z}_{ig}\left[(\vecX_i-\hat{\matm}_g)(\hat{\matPsi}_g^*)^{-1}{{\bf E}_{2ig}^{(2)}}'-\hat{\vecA}_g(\hat{\matPsi}_g^*)^{-1}{{\bf E}_{1ig}^{(2)}}'\right]\left(\sum_{i=1}^Nz_{ig}{\bf E}_{3ig}^{(2)}\right)^{-1}
$$
and
$
\hat{\matsig}_g=\diag(\vecS^L_g),
$
respectively, where
\begin{equation*}\begin{split}
\vecS^L_g=\frac{1}{N_gp}&\sum_{i=1}^N\hat{z}_{ig}\big[b_{ig}(\vecX_i-\hat{\matm}_g)(\hat{\matPsi}_g^*)^{-1}(\vecX_i-\hat{\matm}_g)'-(\hat{\vecA}_g+\hat{\loada}_g{\bf E}_{2ig}^{(2)})(\hat{\matPsi}_g^*)^{-1}(\vecX_i-\hat{\matm}_g)'\\
&-(\vecX_i-\hat{\matm}_g)(\hat{\matPsi}_g^*)^{-1}\hat{\vecA}_g'+a_{ig}\hat{\vecA}_g(\hat{\matPsi}_g^*)^{-1}\hat{\vecA}_g+\hat{\loada}_g{\bf E}_{1ig}^{(1)}(\hat{\matPsi}_g^*)^{-1}\hat{\vecA}_g'\\
&-(\vecX_i-\hat{\matm}_g)(\hat{\matPsi}_g^*)^{-1}{{\bf E}_{2ig}^{(2)}}'\hat{\loada}_g'+\hat{\vecA}_g(\hat{\matPsi}_g^*)^{-1}{{\bf E}_{1ig}^{(2)}}'\hat{\loada}_g'+\hat{\loada}_g{\bf E}_{3ig}^{(2)}\hat{\loada}_g'\big].
\end{split}\end{equation*}

\paragraph{AECM Stage 3} In the third stage, the complete-data consists of the observed data $\vecX_i$, the latent variables ${\bf W}_i$, the labels $\bfz_{i}$ and the latent matrices ${\boldsymbol \fY}^A_i=(\fY^A_{i1},\ldots,\fY^A_{iG})$ for $i=1,\ldots,N$. The complete-data log-likelihood at this stage is
\begin{equation*}\begin{split}
\ell_{\text{C}3}&=C+\sum_{i=1}^N\sum_{g=1}^Gz_{ig}\big[\log \pig+\log h(W_{ig}|\nu_g)+\log \phi_{q\times p}(\fY^A_{ig}|\bzero,W_{ig}\matsigs,\ident_p)\\&
\quad+\log \phi_{n\times p}(\vecX_i|\matm_g+W_{ig}\vecA_g+\fY_{ig}^A\loadb_g',W_{ig}\matsigs,\matPsi_g)\big]\\
&=C+\sum_{i=1}^N\sum_{g=1}^G-\frac{1}{2}z_{ig}\bigg[-n\log|\matPsi_g|+\tr\bigg\{\frac{1}{W_{ig}}\matPsi_g^{-1}(\vecX_i-\matm_g)'\matsigsi(\vecX_i-\matm_g)\\&
\quad-\matPsi_g^{-1}(\vecX_i-\matm_g)'\matsigsi\vecA_g\bigg\}
-\frac{1}{W_{ig}}\matPsi_g^{-1}(\vecX_i-\matm_g)'\matsigsi{\fY_{ig}^A}\loadb_g'-\matPsi_g^{-1}\vecA_g'\matsigsi(\vecX_i-\matm_g)\\
&\quad+W_{ig}\matPsi_g^{-1}\vecA_g'\matsigsi\vecA_g+\matPsi_g^{-1}\vecA_g'\matsigsi{\fY_{ig}^A}\loadb_g'
-\frac{1}{W_{ig}}\matPsi_g^{-1}\loadb_g{\fY_{ig}^A}'\matsigsi(\vecX_i-\matm_g)\\&
\quad+\matPsi_g^{-1}\loadb_g{\fY_{ig}^A}'\matsigsi\vecA_g+\frac{1}{W_{ig}}\matPsi_g^{-1}\loadb_g{\fY_{ig}^A}'\matsigsi{\fY_{ig}^A}\loadb_g'\bigg].
\end{split}\end{equation*}

In the E-step, it can be shown that
\begin{equation*}\begin{split}
\fY_{ig}^A|\vecX_i,&W_{ig},z_{ig}=1\sim\\& \mathn_{n\times r}((\vecX_i-\matm_g-W_{ig}\vecA_g)\matPsi_g^{-1}\loadb_g(\ident_r+\loadb_g'\matPsi_g^{-1}\loadb_g)^{-1},W_{ig}\matsigs,(\ident_r+\loadb_g'\matPsi_g^{-1}\loadb_g)^{-1})
\end{split}\end{equation*}
and so we can calculate the expectations
\begin{equation*}\begin{split}
{\bf E}_{1ig}^{(3)}&\coloneqq\mathbb{E}[\fY_{ig}^A|\hat{\bvtheta},\vecX_i,z_{ig}=1]
=(\vecX_i-\hat{\matm}_g-a_{ig}\hat{\vecA}_g)\matd_g,\\
{\bf E}_{2ig}^{(3)}&\coloneqq\mathbb{E}\left[\frac{1}{W_{ig}}\fY_{ig}^A|\hat{\bvtheta},\vecX_i,z_{ig}=1\right]
=(b_{ig}(\vecX_i-\hat{\matm}_g)-\hat{\vecA}_g)\matd_g,\\
{\bf E}_{3ig}^{(3)}&\coloneqq\mathbb{E}\left[\frac{1}{W_{ig}}{\fY_{ig}^A}'(\matsig_g^*)^{-1}{\fY_{ig}^A}|\hat{\bvtheta},\vecX_i,z_{ig}=1\right]\\
&=n(\ident_r+\hat{\loadb}_g'\hat{\matPsi}_g^{-1}\hat{\loadb}_g)^{-1}+b_{ig}\matd_g'(\vecX_i-\hat{\matm}_g)'\matsigsih(\vecX_i-\hat{\matm}_g)\matd_g\\
&\quad-\matd_g'((\vecX_i-\hat{\matm}_g)'\matsigsih\hat{\vecA}_g+\hat{\vecA}_g'\matsigsih(\vecX_i-\hat{\matm}_g))\matd_g+a_{ig}\matd_g'\hat{\vecA}_g'\matsigsih\hat{\vecA}_g\matd_g,
\end{split}\end{equation*}
where $\matd_g=\hat{\matPsi}_g^{-1}\hat{\loadb}_g(\ident_r+\hat{\loadb}_g'\hat{\matPsi}_g^{-1}\hat{\loadb}_g)^{-1}$.

In the M-step, the updates for $\loadb_g$ and $\matPsi_g$ are calculated. These updates are given by
$$
\hat{\loadb}_g=\sum_{i=1}^N\hat{z}_{ig}[(\vecX_i-\hat{\matm}_g)'\matsigsih{\bf E}_{2ig}^{(3)}-\hat{\vecA}_g'\matsigsih{\bf E}_{1ig}^{(3)}](\sum_{i=1}^Nz_{ig}{\bf E}_{3ig}^{(3)})^{-1}
$$
and
$
\hat{\matPsi}_g=\diag(\vecS^D_g),
$
respectively, where
\begin{equation*}\begin{split}
\vecS^D_g=\frac{1}{N_gp}&\sum_{i=1}^N\hat{z}_{ig}[b_{ig}(\vecX_i-\hat{\matm}_g)'\matsigsih(\vecX_i-\hat{\matm}_g)-(\hat{\vecA}_g'+\hat{\loadb}_g{{\bf E}_{2ig}^{(3)}}')\matsigsih(\vecX_i-\hat{\matm}_g)\\
&-(\vecX_i-\hat{\matm}_g)'\matsigsih\hat{\vecA}_g+a_{ig}\hat{\vecA}_g'\matsigsih\hat{\vecA}_g+\hat{\loadb}_g{{\bf E}_{1ig}^{(3)}}'\matsigsih\hat{\vecA}_g\\
&-(\vecX_i-\hat{\matm}_g)'\matsigsih{\bf E}_{2ig}^{(3)}\hat{\loadb}_g'+\hat{\vecA}_g'\matsigsih{\bf E}_{1ig}^{(3)}\hat{\loadb}_g'+\hat{\loadb}_g{\bf E}_{3ig}^{(3)}\hat{\loadb}_g'].
\end{split}\end{equation*}

Details on initialization, convergence, model selection, and performance criteria are given in Appendix~B. 

\subsection{Reduction in the Number of Free Parameters in the Scale Matrices}
The reduction in the number of free parameters in the scale matrices for each of these models is equivalent to the Gaussian case discussed in \cite{gallaugher18b}.  The reduction in the number of free covariance parameters for the row scale matrix is
\begin{equation}\label{eqn:cond1}
\frac{1}{2}n(n+1)-nq-n+\frac{1}{2}q(q-1)=\frac{1}{2}\big[(n-q)^2-(n+q)\big],
\end{equation}
which is positive for $(n-q)^2>n+q$.
Likewise, for the column scale matrix the reduction in the number of parameters is
\begin{equation}\label{eqn:cond2}
\frac{1}{2}p(p+1)-pr-p+\frac{1}{2}r(r-1)=\frac{1}{2}\big[(p-r)^2-(p+r)\big],
\end{equation}
which is positive for $(p-r)^2>p+r$.

In applications herein, each model is fit for a range of row factors and column factors. If the number of factors chosen by the BIC is the maximum within the range, the number of factors is increased so long as the conditions \eqref{eqn:cond1} and \eqref{eqn:cond2} are met.

\subsection{Semi-Supervised Classification}
Each of the four models presented herein may also be used in the context of semi-supervised classification. Suppose $N$ matrices are observed and $K$ of these observations have known labels from one of $G$ classes. Following \cite{mcnicholas10c}, and without loss of generality, the matrices are ordered so that the first $K$ have known labels and the remaining observations have unknown labels. Using the MMVBFA model for illustration, the observed likelihood is then
\begin{equation*}\begin{split}
L(\bvtheta)=\prod_{i=1}^K\prod_{g=1}^G&\big[\pig\varphi_{n\times p}(\vecX_i~|~\matm_g,\matsig_g+\vecA_g\vecA_g',\matPsi_g+\vecB_g\vecB_g')\big]^{z_{ig}}\\
&\times \prod_{j=K+1}^N\sum_{h=1}^H\pi_h\varphi_{n\times p}(\vecX_i~|~\matm_h,\matsig_h+\vecA_h\vecA_h',\matPsi_h+\vecB_h\vecB_h').
\end{split}\end{equation*}
Whilst it is possible that $H\ne G$, we assume that $H=G$ for the analyses herein. Parameter estimation then proceeds in a similar manner as for the clustering scenario. For more information on semi-supervised classification refer to \cite{mcnicholas10c,mcnicholas16a}.

\subsection{Computational Issues}
One situation that needs to be addressed for all four of these distributions, but particularly the variance-gamma distribution, is the infinite likelihood problem. This occurs as a result of the update for $\hat{\matm}_g$ becoming very close, and in some cases equal to, an observation $\vecX_i$ when the algorithm gets close to convergence. A similar situation occurs in the multivariate case for the mixture of SAL distributions described in \cite{franczak14} and we follow a similar procedure when faced with this issue. While iterating the algorithm, when the likelihood becomes numerically infinite, we set the estimate of $\hat{\matm}_g$ to the previous estimate which we will call $\hat{\matm}^*_g$. We then update $\hat{\vecA}_g$ according to 
$$
\hat{\vecA}^*_g=\frac{\sum_{i=1}^N\hat{z}_{ig}(\vecX_i-{\hat{\matm}}^*_g)}{\sum_{i=1}^N\hat{z}_{ig}a_{ig}}.
$$ 
The updates for all other parameters remain the same. As mentioned in \cite{franczak14}, this solution is a little naive; however, it does generally work quite well.
It is not surprising that this problem is particularly prevalent in the case of the variance-gamma distribution because the SAL distribution arises as a special case of the variance-gamma distribution.

Another computational concern is in the evaluation of the Bessel functions. In the computation of the GIG expected values and the component densities, it may be the case that the argument is far larger than the magnitude of the index---especially in higher dimensional cases.
Therefore, in these situations, the result is computationally equivalent to zero which causes issues with other computations. In such a situation, we calculate the exponentiated version of the Bessel function, i.e., we calculate $\exp(u)K_{\lambda}(u)$ and subsequent calculations can be easily adjusted.

\section{Simulation Study}
A simulation study was performed for each of the four models presented herein. For each of the four models, we consider $d\times d$ matrices with $d\in\{10,30\}$ and, for each value of $d$, we consider datasets coming from a mixture with two components and $\pi_1=\pi_2=0.5$. The datasets have sample sizes $N\in\{100,200,400\}$ and the following parameters are used for all four models for each combination of $d$ and $N$. We take $\matm_1={\bf 0}$ and $\matm_2=\matm_1+\vecC$, where $\vecC$ is a matrix with all entries equal to $c$ for $c\in\{1,2,4\}$. All other parameters are held constant. We take  $\matsig_1=2\ident_d$, $\matsig_2=\ident_d$, $\matPsi_1=\ident_d$, $\matPsi_2=2\ident_d$, and $\vecA_1=\vecA_2={\bf 1}$, where ${\bf 1}$ is a matrix of 1's. Three column factors and two row factors are used with their values being randomly drawn from a uniform distribution on $[-1,1]$. 
See Table~\ref{tab:params} for distribution-specific parameters.

\begin{table}[!htb]
\centering
\caption{Distribution-specific parameters used for the simulations, where the acronyms all take the form MMVDFA and denote ``mixture of matrix variate D factor analyzers'' with $\text{D}\in\{\text{skew-$t$ (ST)},\text{generalized hyperbolic (GH)},\text{variance-gamma (VG)},\text{NIG}\}$.}
{\scriptsize\begin{tabular}{lrr}
\hline
 & Component 1 & Component 2\\
\hline
MMVSTFA & $\nu_1=4$ & $\nu_2=20$\\
MMVGHFA& $\omega_1=4,\lambda_1=-4$ & $\omega_2=10,\lambda_2=4$\\
MMVVGFA& $\gamma_1=4$ & $\gamma_2=10$\\
MMVNIGFA& $\tgamma_1=2$ & $\tgamma_2=4$\\
\hline
\end{tabular}}
\label{tab:params}
\end{table}

We fit the MMVSTFA model to data that is simulated from the MMVSTFA model using the parameters above together with the distribution-specific parameters in Table~\ref{tab:params}. We take an analogous approach with the MMVGHFA, MMVVGFA, and MMVNIGFA models. However, we fit the MMVBFA model to data that is simulated from the MMVVGFA model---this is done to facilitate an illustration that uses data simulated from a mixture of skewed matrix variate distributions.
We fit all models for $G\in\{1,2,3,4\}$ and $q,r\in\{1,2,3,4,5\}$. In Tables~\ref{tab:GQR1} and~\ref{tab:GQR2}, we show the number of times that the BIC correctly chooses the number of groups, row factors, and column factors. In Table~\ref{tab:ARIsim}, the average ARI and corresponding standard deviation for each setting is shown. As one would expect, for each model introduced herein, the classification performance generally improves as $N$ increases. However, this is not the case for the MMVBFA model. In the case $d=10$, it is interesting to note that the number of correct choices made by the BIC for the row and column factors generally decreases as we increase the separation (Table~\ref{tab:GQR1}). However, when $d$ is increased to 30, there is no clear trend in this regard (Table~\ref{tab:GQR2}). The classification performance for the four models introduced here in is excellent overall (Table~\ref{tab:ARIsim}). However, when fitting the MMVBFA model to data simulated from the MMVVGFA model, the BIC never chooses the correct number of groups for $N\in\{200,400\}$. Furthermore, although not apparent from the tables, the model generally overfits the number of groups which, as in the multivariate case, is to be expected when using a Gaussian mixture model in the presence of skewness or outliers.
\begin{table}[!htb]
\centering
\caption{Number of datasets for which the BIC correctly chose the number of groups, row factors, and column factors ($d=10$).}
{\scriptsize\begin{tabular}{ll|ccc|ccc|ccc|ccc|ccc}
\hline
&&\multicolumn{3}{c|}{MMVSTFA}&\multicolumn{3}{|c|}{MMVGHFA}&\multicolumn{3}{|c|}{MMVVGFA}&\multicolumn{3}{|c|}{MMVNIGFA}&\multicolumn{3}{|c}{MMVBFA}\\
$c$&$N$&$G$&$q$&$r$&$G$&$q$&$r$&$G$&$q$&$r$&$G$&$q$&$r$&$G$&$q$&$r$\\
\hline
\multirow{3}{*}{$1$}
&$100$&18&15&19&25&16&24&18&10&12&21&21&20&17&15&19\\
&$200$&23&18&21&25&25&25&21&11&8&24&24&24&0&19&23\\
&$400$&25&21&21&25&25&25&22&17&15&24&24&25&0&19&14\\
\hline
\multirow{3}{*}{$2$}
&$100$&18&14&17&25&9&22&16&7&4&17&18&19&16&17&15\\
&$200$&24&18&19&25&22&22&23&10&2&23&23&24&0&20&20\\
&$400$&25&23&23&25&25&25&19&20&19&25&24&25&0&21&14\\
\hline
\multirow{3}{*}{$4$}
&$100$&8&13&14&23&5&10&17&11&0&24&2&7&19&23&9\\
&$200$&22&9&16&25&4&16&21&8&8&24&7&24&0&18&18\\
&$400$&25&12&21&25&22&12&17&10&19&21&0&14&0&17&19\\
\hline
\end{tabular}}
\label{tab:GQR1}
\end{table}

\begin{table}[!htb]
\centering
\caption{Number of datasets for which the BIC correctly chose the number of groups, row factors, and column factors ($d=30$).}
{\scriptsize\begin{tabular}{ll|ccc|ccc|ccc|ccc|ccc}
\hline
&&\multicolumn{3}{c|}{MMVSTFA}&\multicolumn{3}{|c|}{MMVGHFA}&\multicolumn{3}{|c|}{MMVVGFA}&\multicolumn{3}{|c|}{MMVNIGFA}&\multicolumn{3}{|c}{MMVBFA}\\
$c$&$N$&$G$&$q$&$r$&$G$&$q$&$r$&$G$&$q$&$r$&$G$&$q$&$r$&$G$&$q$&$r$\\
\hline
\multirow{3}{*}{$1$}
&$100$&24&11&12&25&15&18&25&12&12&25&20&21&15&6&2\\
&$200$&25&17&18&25&22&23&25&21&20&25&23&25&0&4&3\\
&$400$&25&22&23&25&25&24&25&25&20&25&25&25&0&10&1\\
\hline
\multirow{3}{*}{$2$}
&$100$&24&15&17&25&17&18&25&13&11&25&22&23&14&5&0\\
&$200$&25&22&19&25&19&22&25&20&22&25&23&25&0&5&3\\
&$400$&25&19&20&25&22&24&25&23&24&25&24&25&0&9&8\\
\hline
\multirow{3}{*}{$4$}
&$100$&24&17&17&25&12&14&25&17&14&25&23&16&18&2&2\\
&$200$&25&18&20&25&18&23&25&21&22&25&21&21&0&3&8\\
&$400$&25&15&15&25&20&24&25&19&20&25&21&22&0&5&7\\
\hline
\end{tabular}}
\label{tab:GQR2}
\end{table}

\begin{table}[!htb]
\centering
\caption{Average ARI values over 25 runs for each setting with standard deviations in parentheses.}
{\scriptsize
\scalebox{0.8}{\begin{tabular}{ll|cc|cc|cc|cc|cc}
\hline
&& \multicolumn{2}{c|}{MMVSTFA}& \multicolumn{2}{c|}{MMVGHFA}& \multicolumn{2}{c|}{MMVVGFA}& \multicolumn{2}{c|}{MMVNIG}& \multicolumn{2}{c}{MMVBFA}\\
$c$&$N$ &$d=10$&$d=30$&$d=10$&$d=30$&$d=10$&$d=30$&$d=10$&$d=30$&$d=10$&$d=30$\\
\hline
\multirow{3}{*}{$1$}&$100$&0.91(0.08)&0.96(0.01)&0.97(0.03)&0.97(0.02)&0.90(0.1)&0.97(0.02)&0.98(0.05)&1.00(0.0)&0.90(0.1)&0.91(0.1)\\
&$200$&0.98 (0.03)&0.99(0.009)&1.00(0.006)&1.00(0.007)&0.97(0.03)&0.99(0.01)&1.00(0.007)&1.00(0.0)&0.75(0.05)&0.76(0.01)\\
&$400$&1.00 (0.005)&1.00(0.004)&1.00(0.0)&1.00(0.0)&0.99(0.03)&1.00(0.0)&1.00(0.006)&1.00(0.0)&0.69(0.1)&0.54(0.07)\\
\hline
\multirow{3}{*}{$2$}
&$100$&0.94(0.03)&0.96(0.02)&0.96(0.03)&0.97(0.03)&0.88(0.1)&0.98(0.02)&0.96(0.07)&1.00(0.0)&0.91(0.1)&0.90(0.1)\\
&$200$&0.98 (0.02)&0.99(0.009)&1.00(0.0)&1.00(0.0)&0.97(0.05)&1.00(0.007)&0.99(0.02)&1.00(0.0)&0.76(0.01)&0.76(0.02)\\
&$400$&1.00 (0.005)&1.00(0.003)&1.00(0.0)&1.00(0.0)&0.98(0.05)&1.00(0.0)&1.00(0.0)&1.00(0.0)&0.66(0.1)&0.53(0.07)\\
\hline
\multirow{3}{*}{$4$}
&$100$&0.84(0.08)&0.97(0.03)&0.94(0.04)&0.97(0.02)&0.92(0.1)&1.00(0.01)&0.98(0.08)&1.00(0.0)&0.94(0.1)&0.93(0.1)\\
&$200$&0.98 (0.02)&1.00(0.004)&1.00(0.0)&1.00(0.0)&0.97(0.05)&1.00(0.0)&0.99(0.02)&1.00(0.0)&0.76(0.02)&0.76(0.01)\\
&$400$&1.00 (0.004)&1.00(0.002)&1.00(0.0)&1.00(0.0)&0.98(0.03)&1.00(0.0)&1.00(0.03)&1.00(0.0)&0.73(0.08)&0.54(0.07)\\
\hline
\end{tabular}}}
\label{tab:ARIsim}
\end{table}

\section{MNIST Digits}
\cite{gallaugher18a,gallaugher18b} consider the MNIST digits dataset; specifically, looking at digits 1 and 7 because they are similar in appearance. Herein, we consider the digits 1, 6, and 7. This dataset consists of 60,000 (training) images of Arabic numerals 0 to 9. We consider different levels of supervision and perform either clustering or semi-supervised classification. Specifically we look at 0\% (clustering), 25\%, and 50\% supervision. For each level of supervision, 25 datasets consisting of 200 images each of digits 1, 6, and 7 are taken. As discussed in \cite{gallaugher18a}, because of the lack of variability in the outlying rows and columns of the data matrices, random noise is added to ensure non-singularity of the scale matrices. Each of the four models developed herein, as well as the MMVBFA model, are fitted for 1 to 17 row and column factors. In Table~\ref{tab:ARIMNIST}, the average ARI and misclassification rate (MCR) values are presented for each model and each level of supervision.
\begin{table}[!htb]
\centering
\caption{Average ARI and MCR values for the MNIST dataset for each level of supervision, with respective standard deviations in parentheses for digits 1,6, and 7.}
{\scriptsize\begin{tabular}{ll|rrrrr}
\hline
Supervision&&MMVSTFA&MMVGHFA&MMVVGFA&MMVNIGFA&MMVBFA\\
\hline
\multirow{2}{*}{0\% (clustering)}&${\text{ARI}}$&0.58(0.09)&0.58(0.09)&0.62(0.1)&0.47(0.1)&0.36(0.09)\\
&${\text{MCR}}$&0.17(0.04)&0.17(0.08)&0.15(0.04)&0.22(0.05)&0.28(0.09)\\
\hline
\multirow{2}{*}{25\%}&${\text{ARI}}$&0.72(0.1)&0.72(0.1)&0.75(0.1)&0.64(0.2)&0.51(0.16)\\
&${\text{MCR}}$&0.10(0.04)&0.10(0.04)&0.094(0.04)&0.14(0.07)&0.20(0.07)\\
\hline
\multirow{2}{*}{50\%}&${\text{ARI}}$&0.83(0.07)&0.85(0.03)&0.83(0.07)&0.81(0.1)&0.72(0.06)\\
&${\text{MCR}}$&0.059(0.03)&0.052(0.02)&0.061(0.03)&0.067(0.05)&0.10(0.06)\\
\hline
\end{tabular}}
\label{tab:ARIMNIST}
\end{table}

In the completely unsupervised case, three of the skewed models have a MCR of around 16\%. However, at 25\% supervision, this decreases to around 10\% and, at 50\% supervision, this falls again to around 5\%.  At all three levels of supervision, it is clear that all four skewed mixture models introduced herein outperform the MMVBFA model. In fact, the performance of the MMVBFA model at 25\% supervision is not as good as that of the MMVVGFA, MMVGHFA or MMVSTFA  models in the completely unsupervised case (i.e., 0\% supervision). 

\begin{figure}[!htb]
\centering
\includegraphics[width=0.5\textwidth,height=0.5\textwidth]{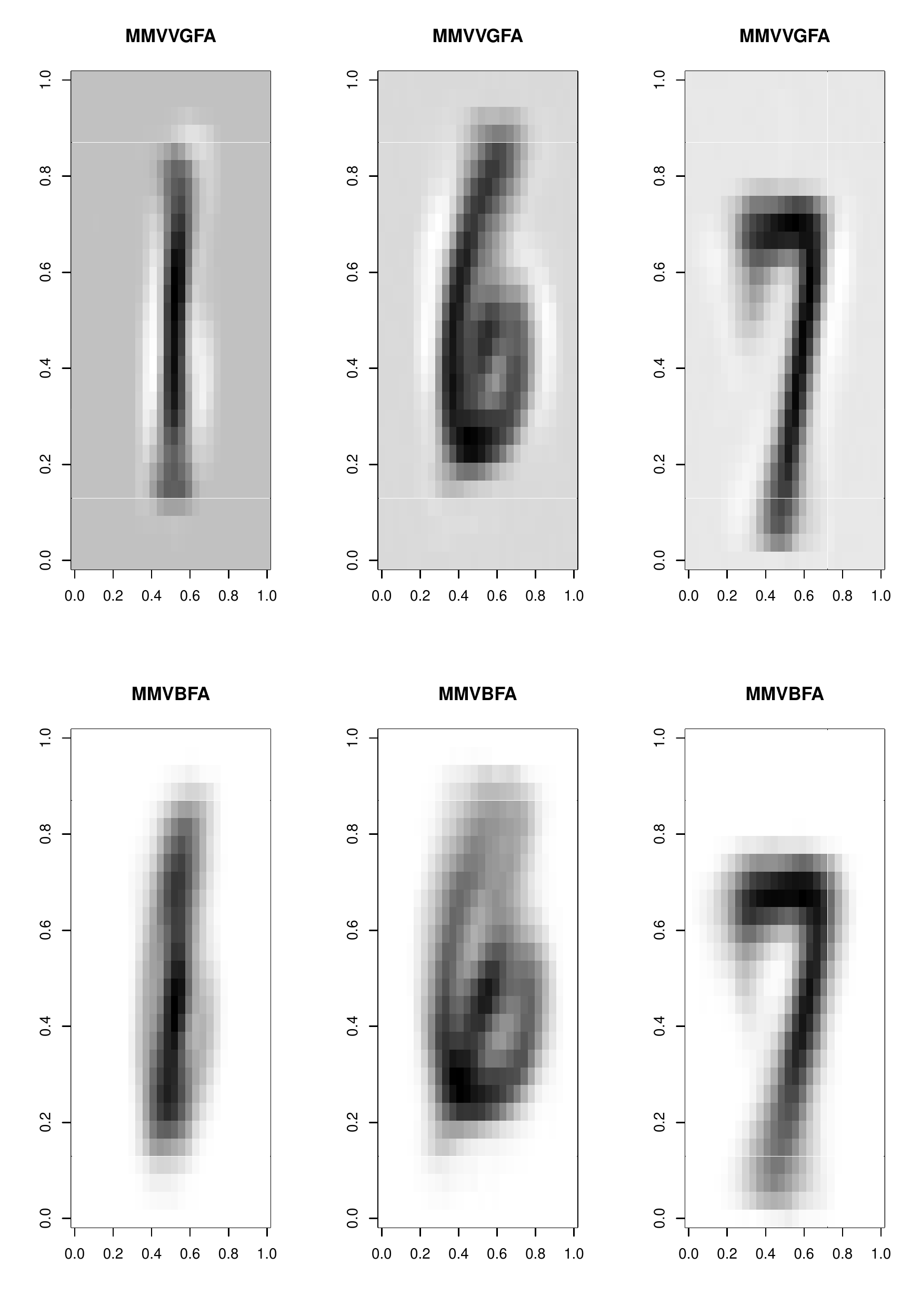}
\caption{Heat maps of estimated location matrices for the MMVBFA and MMVVGFA models for each class in the unsupervised case.}
\label{fig:MNIST}
\end{figure}

It is of interest to compare heat maps of the estimated location matrices for the MMVBFA and MMVVGFA models for one of the datasets in the unsupervised case (Figure~\ref{fig:MNIST}). It can be seen that the images are a lot clearer for the MMVVGFA model compared to the MMVBFA model. This is particularly prominent when considering the the results for the digit 6, for which one can see a possible 1 or 7 in the background for the MMVBFA heat map. Moreover, for digit 1, one can see a faint 6 in the background when looking at the MMVBFA heat map.

\section{Discussion}
The MMVBFA model has been extended to four skewed distributions; specifically, the matrix variate skew-$t$, generalized hyperbolic, variance-gamma, and NIG distributions. AECM algorithms were developed for parameter estimation, and the novel approaches were illustrated on real and simulated data. In the simulations, the models introduced herein generally exhibited very good performance under various scenarios. As expected, the MMVBFA model did not perform well when applied to data from the MMVVGFA model.
In the real data example, all four of the skewed matrix variate models introduced herein performed better than the MMVBFA model. As one would expect, the difference in performance was most stark in the clustering case. 
 
Software to implement the approaches introduced herein, written in the Julia language \citep{bezanson17,mcnicholas19}, is available in the {\tt MatrixVariate.jl} repository \citep{pocuca19}. Future work will include considering a family of models similar to the parsimonious Gaussian mixture models of \cite{mcnicholas08,mcnicholas10d}. Another area of future work would be to compare this method of directly modelling skewness to using transformations such as those found in \cite{melnykov18, melnykov19}. It might also be of interest to consider matrix variate data of mixed type, which would allow, for example, analysis of multivariate longitudinal data where the variables are of mixed type. Finally, this methodology could be extended to mixtures of multidimensional arrays \citep[see][]{tait19}, which would be useful for studying several data types including coloured images and black and white movie clips.

\section*{Acknowledgements}

This work was supported by a Vanier Canada Graduate Scholarship (Gallaugher), the Canada Research Chairs program (McNicholas), and an E.W.R.\ Steacie Memorial Fellowship (McNicholas).

{\small

\appendix

\section{Mixture of Factor Analyzers Model}\label{app:mfa}
Because of data becoming increasingly higher dimensional, dimension reduction techniques are becoming ever more important. In the multivariate case, the mixture of factor analyzers model is widely used. Reverting back to the notation where $\vecX_i$ represents a $p$-dimensional random vector, with $\bfx_i$ as its realization, the factor analysis model for $\vecX_1,\ldots,\vecX_n$ is
$$
\vecX_i=\vecmu+\bLambda\vecU_i+\vecepsilon_i,
$$
where $\vecmu$ is a location vector, $\bLambda$ is a $p\times q$ matrix of factor loadings with $q<p$, $\vecU_i\sim\mathcal{N}_q({\bf 0},\ident)$ denotes the latent factors, $\vecepsilon_i\sim \mathcal{N}_q({\bf 0},\matPsi)$, where $\matPsi=\text{diag}(\psi_1,\psi_2,\ldots,\psi_p)$, and $\vecU_i$ and $\vecepsilon_i$ are each independently distributed and independent of one another. Under this model, the marginal distribution of $\vecX_i$ is $\mathcal{N}_p(\vecmu,\bLambda\bLambda'+\matPsi)$. Probabilistic principal component analysis (PPCA) arises as a special case with the isotropic constraint $\matPsi=\psi\ident_p$ \citep{tipping99a}.

\cite{ghahramani97} develop the mixture of factor analyzers model, which is a Gaussian mixture model with covariance structure $\matsig_g=\bLambda_g\bLambda_g'+\matPsi$. A small extension was presented by \cite{mclachlan00a}, who utilize the more general structure $\matsig_g=\bLambda_g\bLambda_g'+\matPsi_g$. \cite{tipping99b} introduce the closely-related mixture of PPCAs with $\matsig_g=\bLambda_g\bLambda_g'+\psi_g\ident$. \cite{mcnicholas08} constructed a family of eight parsimonious Gaussian models by considering combinations of the constraints $\loada_g=\loada$, $\matPsi_g=\matPsi$ and $\matPsi_g=\psi_g\ident$. \cite{mcnicholas10d} and \cite{bhattacharya14} extend the work \cite{mcnicholas08}. There has also been work on extending the mixture of factor analyzers to other distributions, such as the skew-$t$ distribution \citep{murray14b,murray17a}, the generalized hyperbolic distribution \citep{tortora16}, the skew-normal distribution \citep{lin16}, the variance-gamma distribution \citep{smcnicholas17}, and others \citep[e.g.,][]{murray17a}.

\section{Model Selection, Convergence, Performance Evaluation Criteria, and Initialization}
In general, the number of components, row factors, and column factors are not known {\it a priori} and therefore need to be selected. In our simulations and analyses, the Bayesian information criterion \cite[BIC;][]{schwarz78} is used. The BIC is given by
$$
\text{BIC}=2\ell(\hat{\bvtheta})-\rho\log N,
$$
where $\ell(\hat{\bvtheta})$ is the maximized log-likelihood and $\rho$ is the number of free parameters.

A simple convergence criterion is based on lack of progress in the log-likelihood, where the algorithm is terminated when $l^{(t+1)}-l^{(t)}<\epsilon$, where $\epsilon>0$ is a small number. Oftentimes, however, the likelihood can plateau before increasing again, thus using lack of progress would terminate the algorithm prematurely \citep[see][for examples]{mcnicholas10a}. Another option, and one that is used for our analyses, is a criterion based on the Aitken acceleration \citep{aitken26}. The Aitken acceleration at iteration $t$ is
$$
a^{(t)}=\frac{l^{(t+1)}-l^{(t)}}{l^{(t)}-l^{(t-1)}},
$$
where $l^{(t)}$ is the observed likelihood at iteration $t$. We then have an estimate, at iteration $t+1$, of the log-likelihood after many iterations:
$$
l_{\infty}^{(t+1)}=l^{(t)}+\frac{(l^{(t+1)}-l^{(t)})}{1-a^{(t)}}
$$
\citep{bohning94,lindsay95}. As suggested by \cite{mcnicholas10a}, the algorithm is terminated when $l_{\infty}^{(k+1)}-l^{(k)}\in(0,\epsilon)$. It should be noted that we set the value of $\epsilon$ based on the magnitude of the log-likelihood. Specifically, for each AECM algorithm in our analyses, after five iterations we set $\epsilon$ to a value three orders of magnitude lower than the log-likelihood.

To assess classification performance, the adjusted Rand index \cite[ARI;][]{hubert85} is used. The ARI is the Rand index \citep{rand71} corrected for chance agreement. The ARI compares two different partitions---in our case, predicted and true classifications---and takes a value of 1 if there is perfect agreement. The expected value of the ARI under random classification is 0. 

Finally, there is the issue of initialization. In our simulations and data analyses, we used soft initializations by generating group memberships at random using a uniform distribution. From these initial soft group memberships $\hat{z}_{ig}$, we initialize the location matrices using 
$$
\hat{\matm}_g=\frac{1}{N_g}\sum_{i=1}^N{\hat{z}_{ig}\vecX_i},
$$ where $N_g=\sum_{i=1}^N\hat{z}_{ig}$.
Each skewness matrix is initialized as a matrix with all entries equal to 0.1---note that matrices with all entries equal to 0 cannot be used because the component densities would not be defined. The diagonal scale matrices, $\matsig_g$ and $\matPsi_g$ are initialized as follows
$$
\hat{\matsig}_g=\frac{1}{pN_g}\diag\left\{{\sum_{i=1}^N\hat{z}_{ig}(\vecX_i-\hat{\matm}_g)(\vecX_i-\hat{\matm}_g)'}\right\},
$$
and 
$$
\hat{\matPsi}_g=\frac{1}{nN_g}\diag\left\{{\sum_{i=1}^N\hat{z}_{ig}(\vecX_i-\hat{\matm}_g)'(\vecX_i-\hat{\matm}_g)}\right\}.
$$
The factor loadings are initialized randomly from a uniform distribution on $[-1,1]$.

}

\end{document}